# Lanthanide L-Edge Spectroscopy of High-Entropy Oxides: Insights into Valence and Phase Stability

Gerald R. Bejger[1], Mary Kathleen Caucci[2], Saeed S.I. Almishal[3], Billy Yang[3], Jon-Paul Maria[3], Susan B. Sinnott[3], Christina M. Rost[1,*]

1. Department of Materials Science and Engineering, Virginia Polytechnic Institute and State University, Blacksburg, VA 24060, USA.
2. Department of Chemistry, The Pennsylvania State University, University Park, PA 16802, USA.
3. Department of Materials Science and Engineering, The Pennsylvania State University, University Park, PA 16802, USA.
4. Materials Research Institute, The Pennsylvania State University, University Park, PA 16802, USA
5. Institute for Computational and Data Science, The Pennsylvania State University, University Park, PA 16802, USA

*Corresponding author: cmrost@vt.edu

**Abstract**

High-entropy oxides (HEOs) are a promising class of multicomponent ceramics with tunable structural and electronic properties. In this study, we investigate the local electronic structure of rare-earth HEOs in the (Ce, Sm, Pr, La, Y)O$_2$ system using X-ray absorption spectroscopy (XAS). By systematically increasing the Ce concentration, we observe a phase transition from bixbyite to fluorite, tracked by X-ray diffraction (XRD) and corroborated by L-edge XANES analysis of La, Sm, Ce, and Pr. The oxidation states of La and Sm remain trivalent, while Ce exhibits a minor Ce$^{3+}$ fraction and Pr shows a consistent mixed-valence state. Density functional theory (DFT) calculations with Bader charge analysis support these findings and reveal that the phase transition is driven by compositional effects rather than cation redox. Our combined experimental and computational approach provides new insights into structure–valence correlations in RE-HEOs and their implications for ionic transport and phase stability.

## Introduction

Harnessing compositional complexity offers an unprecedented route to tailor electronic, ionic, and thermal functionalities for next-generation energy, catalysis, and electronic materials [1]. High-entropy materials, first conceptualized independently by Yeh [2] and Cantor [3] in 2004, incorporate five or more principal cations into a crystalline lattice, enabling the formation of unique compositional phases and uncharted space over unique material properties. This concept was later extended to oxides in 2015 by Rost et al. [4], who demonstrated the stabilization of what would otherwise be a thermodynamically unfavorable mixture of oxide components. The resulting high entropy oxides exhibit unique electronic [5], [6], ionic [7], and thermal properties [8], positioning them as promising candidates for a wide range of advanced applications. The compositional complexity of these materials can influence the electronic structure due to the cation diversity which in turn introduces varying orbital energies. This elemental diversity could result in a narrowing or broadening of the band gap.

Rare earth oxides commonly adopt fluorite or bixbyite structures, with CeO$_2$ as a prototypical fluorite-phase oxide. Fluorite rare earth oxides support high oxygen ionic conductivity due to their ability to form oxygen vacancies, especially when doped with aliovalent cations like Gd$^{3+}$ or Sm$^{3+}$[9]. Ce and Pr exhibit mixed valence (Ce$^{4+}$/Ce$^{3+}$, Pr$^{3+}$/Pr$^{4+}$), enabling redox activity that supports catalytic and electronic transport applications [10]. Reduced $CeO_{2-\delta}$ supports polaron hopping between $Ce^{3+}$ and $Ce^{4+}$, enabling memristive behavior, particularly in thin-film devices where electric fields rearrange vacancies to form or disrupt conductive paths [11], [12]. The rare-earth (RE) sesquioxide HEO system, $(CeSmPrLaY)O_{2-\delta}$, initially reported by Djenadic et al. [13], offers an intriguing platform for exploring how cation diversity influences ionic and electronic structures. Their work demonstrates that a single-phase high entropy oxide could be synthesized despite the differences in cation size and electronic configuration, with the material adopting what they attribute to a bixbyite (Ia-3) structure. Using this composition, Kotsonis demonstrated resistive switching behavior as part of his doctoral thesis [14]. In a related study, Sarkar et al. [15] synthesized equiatomic rare earth oxides with up to seven different cations using nebulized spray pyrolysis and found that $Ce^{4+}$ and multivalent Pr played key roles in stabilizing a fluorite (Fm-3m) structure and narrowing the band gap to 1.95–2.14 eV. Their work emphasizes the importance of redox-active species and oxygen vacancies in tuning structural and electronic properties. Building on this foundation, Riley et al. [16] in 2021 systematically tuned the composition of this RE-HEO by increasing the concentration of Ce to 20, 50, 80, and 100% using a sol gel route. While the sol-gel approach yielded single phase materials for each compositional variant, the solid-state method was not a single phase at 20% Ce. This behavior reflects the interplay between synthesis method, composition, and phase stability, as the bixbyite structure transitions into a fluorite (Fm-3m) structure at 50% Ce concentration.

The stability of bixbyite and fluorite crystal structures is strongly influenced by the valence state, preferred coordination environment, ionic size of the constituent cations, and thermal history [17]. Fluorite (Fm-3m) favors cations with higher oxidation states (typically 4+) and eightfold coordination [18], while bixbyite including, type C Ia-3 stabilizes lower valence (3+) cations in distorted sixfold coordination [19]. As the average cation charge increases, as in Ce-rich compositions, the fluorite phase becomes thermodynamically preferred at high enough temperatures, whereas 3+ dominated rare-earth-rich compositions favor bixbyite [20], [21]. Size mismatch also plays a critical role: fluorite tolerates more size disparity among cations than bixbyite [22]. These crystal chemical trends manifest clearly in phase field maps, where increasing Ce concentration drives a transition from bixbyite to fluorite structure. Solubility rules similarly reflect these factors, with high solubility achieved when cation charge and radius are closely matched to the host lattice. Understanding the interplay of these parameters is essential for predicting phase stability and tailoring the synthesis of multicomponent oxides. The bixbyite structure can be described as a derivative of the fluorite structure, where the lattice parameter is doubled, and with more unoccupied anion sites relative to the fluorite structure [23]. The fluorite structure is known for increased oxygen ionic conductivity when doped with lower valence elements such as Sm, the ionic conductivity increases substantially [9].

X-ray photoelectron spectroscopy (XPS) from both Djenadic and Riley's work suggested that Ce occupies an oxidation state of nearly 4+, while Pr occupied a mixed state between 3+ and 4+ in the equimolar sample. Due to the overlap of 3+ and 4+ peaks in XPS for both Ce and Pr, deconvolution is difficult. Despite the importance of accurate electronic and structural characterization, much of the current understanding relies on limited or surface-sensitive techniques for resolving these complexities. To uncover the cation environment and structural nature of these high-entropy oxides, more powerful, element-specific probes, such as X-ray absorption measurements, are necessary.

In this work, we investigate the valence state and coordination environment of the elements in $Ce_x(SmPrLaY)_{1-x}O_{2-\delta}$, where Ce content was increased as x = 20, 32.5, and 40%, with other elements maintained in an equimolar amount using X-ray absorption fine structure spectroscopy (XAFS). This element-specific technique is promising for HEO analysis, as it probes the local electronic structure of atoms in the sample, providing information on cation charge state and coordination environment through analysis of the near-edge fine structure (XANES).

Increasing the Ce concentration in $Ce_x(SmPrLaY)_{1-x}O_{2-\delta}$, where x = 20, 32.5 and 40%, results in a phase transition from the bixbyite to fluorite structure. To investigate this transition, we performed XAFS measurements of the La, Ce, Pr, and Sm $L_3$ absorption edges in the series $Ce_x(SmPrLaY)_{1-x}O_{2-\delta}$ (x = 20, 32.5 and 40%) and selected standards. This region of the XAFS provides invaluable information on the bonding environment and valence state of the measured elements. La and Sm are mainly trivalent elements, so we investigated their absorption edges for coordination number, while we examined Ce and Pr for oxidation state. To further explore the electronic structure and oxidation states of the cations, density functional theory (DFT) calculations were employed, offering a complementary theoretical framework to the experimental observations.

**Experimental Methods**

High-entropy oxide samples were synthesized via a conventional solid-state reaction method. Stoichiometric amounts of the precursor oxides; $La_2O_3$ (99.99%), $CeO_2$ (99.99%), $Sm_2O_3$ (99.99%), $Pr_6O_{11}$ (99.99%), and $Y_2O_3$ (99.99%) were mixed to achieve the desired composition. The mixed powders were then milled in methanol in a vibratory mill with Y-stabilized $ZrO_2$ (YSZ) milling media (2 mm, 3 mm, and 5 mm in near equal proportions) to promote homogeneity and reduce particle size. Following milling, the powders were compacted into 1.27 cm diameter pellets and reactively sintered at 1400 °C for 48 hours in air to promote phase formation and crystallization. After sintering, the samples were air-quenched to preserve the high-temperature phase. These pellets were characterized via X-ray diffraction (XRD) before and after being ground in an agate mortar and pestle for 10 minutes. The ground powders were then the subject of this work. For comparison and calibration, standard compounds with known oxidation states and coordination environments were synthesized according to the procedures detailed in the Supplementary Information (Table S1). These standards were used for XAFS spectra analysis to aid in determining La and Sm coordination trends and the oxidation states of Ce and Pr in the HEO samples.

X-ray diffraction (XRD) measurements were performed using a Malvern Panalytical Empyrean (Almelo, Netherlands) with Cu $K_\alpha$ radiation and operating at 45 kV and 40 mA equipped with iCore and dCore prefix modules. The incident optics included a 14 mm mask, and a divergence slit angle of 1/2° while the diffracted beam optics included a 1/2° anti-scatter slit and 0.04 rad soller slit. X-ray absorption fine structure measurements were performed on beamline 12-BM at the Advanced Photon Source, Argonne National Laboratory (Lemont, IL). XAFS measurements were collected in fluorescence and transmission mode simultaneously, the spectra with the best signal to noise ratio was used in this work. The fluorescence detector used a Hitachi Vortex-ME7 silicon drift detector, while transmission mode data was collected via ionization chambers. The XAFS samples were prepped for transmission mode, with the sample mass being determined from xraydb [24]. The XAFS data was analyzed using a combination of the Demeter package for XAFS analysis [25] and Larch [26].

**Computational Methods**

First-principles DFT calculations were carried out using VASP (version 6.3.0) [27], [28], employing the projector augmented-wave (PAW) method [29]. The r$^2$SCAN meta-GGA functional [30] was used to treat exchange-correlation effects, as it has demonstrated strong performance for rare-earth oxide systems [31]. A kinetic energy cutoff of 700 eV was applied for the plane-wave basis, and electronic self-consistency was achieved with an energy criterion of 10$^{-6}$ eV. Ionic relaxations proceeded until all atomic forces were below 0.02 eV/Å. Structural optimization used the conjugate gradient algorithm. Γ-centered k-point meshes of 2×2×2 were constructed using a k-spacing of 0.4 Å$^{-1}$. The pseudopotentials Ce, La, Pr, Sm, Y_sv, and O were selected from the PAW 64 dataset, with the f electrons of Ce, Pr, and Sm explicitly treated as valence electrons. All calculations assumed initial ferromagnetic spin alignment.

To evaluate atomic charges, Bader charge analysis was performed. The program developed by Henkelman and co-workers [32], [33], [34] was employed to partition space into Bader volumes based on charge density maxima. Valence charge densities were integrated within these regions to obtain the Bader charges. The analysis utilized the approximate all-electron charge density derived from combining the AECCAR and CHGCAR files generated by VASP. Net ionic charges were calculated by subtracting the integrated Bader valence charges from the nominal valence electron counts defined in the PAW pseudopotentials.

To model the disordered oxide compositions, seven 96-atom fluorite supercells ($Fm\overline{3}m$) were constructed using 2×2×2 conventional unit cells of Ce$_x$(SmPrLaY)$_{1-x}$O$_{2-\delta}$ with Ce concentrations x = 0.22, 0.25, 0.31, 0.34, 0.38, 0.50, and 0.81. The 32-site cation sublattices in each supercell were populated using the special quasi-random structure (SQS) approach [35], implemented via the Integrated Cluster Expansion Toolkit (ICET) [36]. Identical cation configurations were applied to generate the corresponding bixbyite ($Ia\overline{3}$) supercells, which also contain 32 cations per conventional unit cell. For each composition, several oxygen vacancy concentrations (δ) were examined and tabulated in Table S2. In fluorite structures, oxygen vacancies were introduced randomly. In contrast, oxygen vacancies in bixbyite structures were restricted to the Wyckoff *8b*

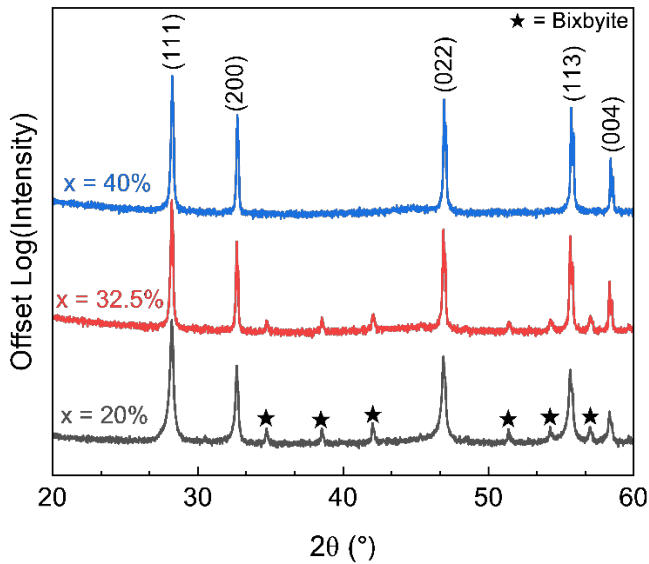

Figure 1: X-ray diffraction results of Ce$_x$(SmPrLaY)$_{1-x}$O$_{2-\delta}$, shows that increasing the Ce content of the composition to 40% results in a single-phase fluorite structure. The black stars denote peaks related to the bixbyite, Ia-3 symmetry. The main fluorite Fm-3m peak indices are indexed at the top of the figure.

positions, in accordance with the crystallographic site symmetry.

## Results and Discussion

XRD results for the Ce concentration series of $Ce_x(SmPrLaY)_{1-x}O_2$ are shown in Figure 1. The diffraction patterns reveal a clear structural evolution as the Ce content increases. At 20% Ce concentration, the diffraction pattern exhibits characteristic peaks corresponding to the bixbyite (Ia-3) phase, similarly to [13], [16]. As the Ce content increases to 32.5%, bixbyite-related peaks weaken, suggesting a transition toward the fluorite (Fm-3m) phase. By 40% Ce concentration, the bixbyite peaks disappear entirely, and the diffraction pattern aligns fully with that of the fluorite phase, indicating a complete transformation. The disappearance of bixbyite-related peaks and the stabilization of fluorite symmetry confirm that increasing Ce content plays a crucial role in driving the structural transition from bixbyite to fluorite.

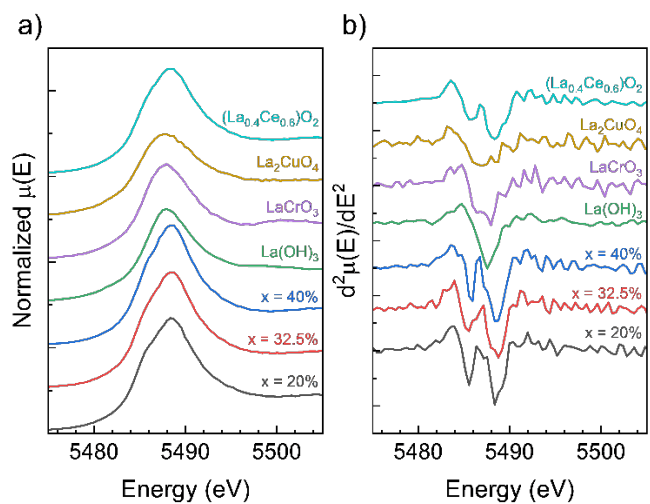

Figure 2: La $L_3$ absorption edge (a) and second derivative (b) in $Ce_x(SmPrLaY)_{1-x}O_{2-\delta}$ and measured standards.

**La Absorption Edge:** XAFS results for the lanthanum (La) $L_3$ absorption edges, and their second derivatives, of the Ce concentration series and measured standards are shown in Figure 2a and 2b, respectively. The $L_3$ absorption edge of La mainly corresponds to the excitation of a 2p core electron to unoccupied d states. The edge position of the XANES region can indicate oxidation state, and it is seen that the edge position of La in the HEO series is similar to that of the measured 3+ standards. This can be visualized by using the maximum of the first derivative, shown in Figure S1. $La^{3+}$ has an electron configuration with no f electrons, therefore the shape of the white line in the XANES spectra should reflect the empty 5d orbital and can be used to investigate local environment. The shape of the white line has distinct features that can hint at coordination environment. As seen in the RE-HEOs and the La doped Ce standard, a shoulder is present on the white line. We attribute this feature to the ligand-field splitting of the 5d orbital [37]. The second derivatives of the absorption spectra show 2 peaks, which supports this reasoning. The shorter shoulder being on the pre-edge is consistent with expected d orbital splitting for a fluorite-like structure, where the $e_g$ band is at a lower energy level than the $t_{2g}$ band [38]. Based off the first minimum in the second derivative shown in Figure 2b, the splitting energy remains consistent throughout the addition of Ce, suggesting a consistent coordination environment.

**Sm Absorption Edge:** Figure 3a and 3b show the Sm $L_3$ and Sm $L_1$ absorption edges, respectively. Based on the edge energy, Sm also maintains a 3+ oxidation state throughout the increasing Ce concentration series. However, the splitting of the white line is not observed, as the presence of f electrons implies stronger electron shielding, and the ionic radius of $Sm^{3+}$ is less than that of $La^{3+}$, which would lead to less d orbital splitting [39], [40]. Instead of noticeable d-orbital splitting as seen on the La $L_3$ absorption edge, the Sm $L_3$ white line for all measured samples is a single broad, near symmetric peak. This peak is attributed to an electric dipole transition from 2p to 5d states [41]. The combination of the $L_3$ and $L_1$ Sm absorption edges can be fit to provide insight into the potential local environment. This is because the $L_3$ edge exhibits broadening as the width is associated with d band broadening [41], [42]. The pre-edge feature on the $L_1$ edge is attributed to an electronic transition generated by the hybridization of p and d states and the degree of

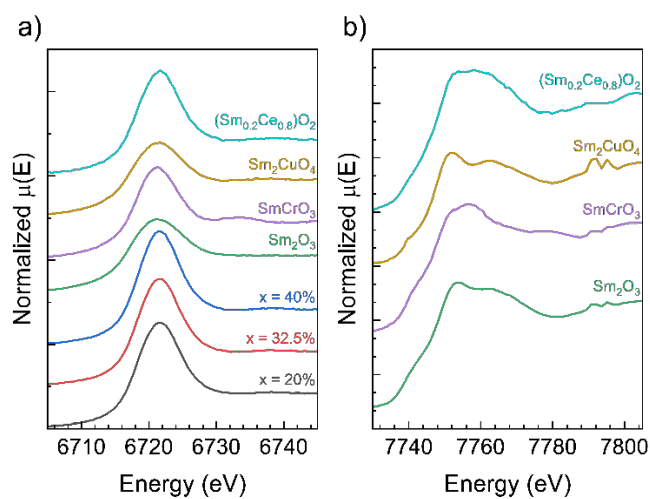

Figure 3: (a), (b): Sm $L_3$ and Sm $L_1$, respectively, absorption edge of Sm in $Ce_x(SmPrLaY)_{1-x}O_{2-\delta}$ and measured standards

hybridization should reflect the first coordination shell disorder around the Sm cation [41]. A fitting procedure, similar to what was done by Asakura et al. [41], was performed to gain insight into the local environment of Sm. For the $L_3$ edge, an arctangent step function was used to represent the excitation to continuum, while a single pseudo-Voight function was used to fit the white line. For the $L_1$ edge, the pre-edge feature was fit with a gaussian and a linear plus Lorentzian function. Examples of the fitting procedure are shown in Figure S2. The Sm $L_1$ edge of the HEO series had weak intensity due to the low concentration of Sm and absorption effects of the other cations in the material. Because of this, only the full width half max (FWHM) of the Sm $L_3$ edge of the HEO samples were fit and plotted on a line of best fit to gain insight to the potential local environment of Sm. Figure 4 illustrates the relationship between the Sm $L_3$ FWHM and the Sm $L_1$ peak area for a series of Sm-containing oxides, including the HEO samples and standard reference compounds. The trend observed in the standards suggests a correlation between the broadening of the Sm $L_3$ edge and the Sm $L_1$ peak area, as seen by

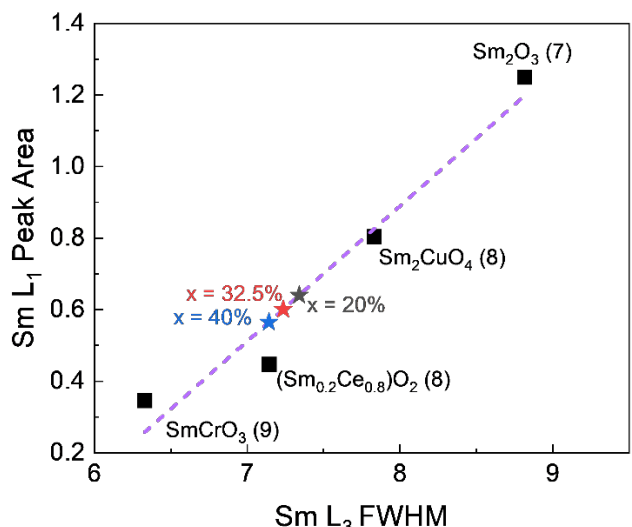

Figure 4: Correlation between the pre-edge peak area of Sm $L_1$ edge XANES spectra and the FWHM of the white line of Sm $L_3$ edge XANES. HEO samples are labeled as stars while standards are labeled as squares.

Asakura [41]. Notably, the $L_3$ FWHM of the 40% Ce-substituted sample aligns closely with $(Sm_{0.2}Ce_{0.8})O_2$, a fluorite-structured standard. This suggests that at higher Ce concentrations, the Sm environment in the HEO system adopts a local electronic structure characteristic of the fluorite phase. This supports the XRD findings that indicate a fluorite-dominant structure at 40% Ce. In contrast, the lower Ce concentration samples (20% and 32.5%) deviate slightly from this trend, implying a more disordered environment, consistent with the presence of bixbyite-like features observed in XRD.

**Ce Absorption Edge:** The Ce $L_3$ edge energy of the HEO samples closely aligns with that of the

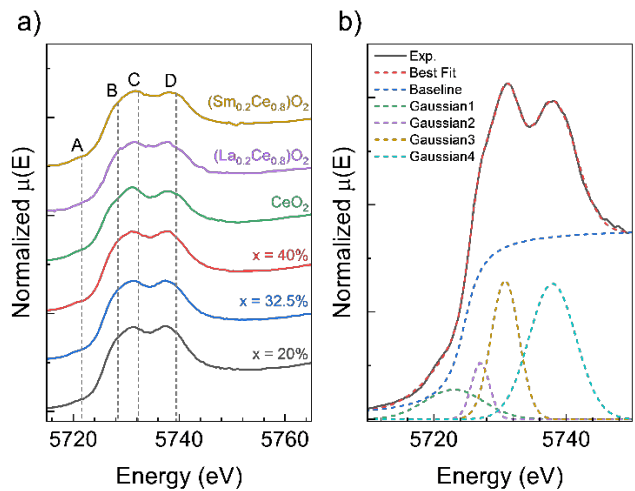

Figure 5: a) Ce $L_3$ absorption edge of Ce in $Ce_x(SmPrLaY)_{1-x}O_{2-\delta}$ and measured standards, peaks used for fitting labeled A-D and are defined in the text. b) Example of fitting procedure for the Ce $L_3$ edge on $CeO_2$.

$CeO_2$ standard, which is primarily $Ce^{4+}$. However, $CeO_2$ is known to contain trace amounts of $Ce^{3+}$ due to the presence of oxygen vacancies, leading to a small contribution from reduced cerium species [43]. This suggests that the Ce oxidation state in the HEO may also include a minor $Ce^{3+}$ component, making a more detailed spectral analysis to quantify its contribution necessary. To estimate the average oxidation state of Ce as its concentration increases, Ce $L_3$ absorption edge fitting was performed using a series of four Gaussian functions, following the method outlined in [44]. An example of this fitting is shown in Figure 5b. These four Gaussians correspond to distinct electronic transitions in the Ce $L_3$ edge spectrum and are labeled as peaks A, B, C, and D in Figure 5a. Peak A is a pre-edge feature that is attributed to final states with delocalized d character. Peak B is associated with Ce 3+ and is assigned to the Ce transition from 2p to $4f^15d$. Peaks C and D are associated with $Ce^{4+}$ final states of $2p4f^05d$ and $2p4f^15dL$, respectively, where L denotes an oxygen ligand 2p hole [44]. Fitting these peaks and taking a ratio of peak area, it is possible to estimate the percentage of $Ce^{3+}$ that is present in the RE-HEO series. The results, shown in Table 1, indicate that Ce maintains a mixed valence state across all measured compositions, with approximately 10% $Ce^{3+}$ present consistently, regardless of total Ce concentration. This observation is in agreement with prior X-ray photoelectron spectroscopy (XPS) studies on $(LaCePrSmY)O_2$, which also found evidence of a small but stable $Ce^{3+}$ fraction [16].

Table 1: Valence estimation for Ce and Pr in $Ce_x(SmPrLaY)_{1-x}O_{2-\delta}$. The $Ce^{4+}$ percentage was calculated by subtracting the measured $Ce^{3+}$% from 100, then multiplying by the Ce mole fraction (x = 0.2, 0.325, 0.4) to get %$Ce^{4+}$ added. This approach shows a consistent $Ce^{3+}$ percentage and Pr average oxidation state across increasing $Ce^{4+}$ content, indicating a stable redox behavior.

| Sample | $CeO_2$ | x = 20% | x = 32.5% | x = 40% |
| --- | --- | --- | --- | --- |
| % $Ce^{3+}$ | 9.2 ± 3.0 | 10.9 ± 3.1 | 9.6 ± 3.4 | 11.7 ± 3.6 |
| % $Ce^{4+}$ added | -- | 17.82 | 29.38 | 35.32 |
| Pr estimated valence | -- | 3.60 ± 0.09 | 3.54 ± 0.1 | 3.56 ± 0.09 |

**Pr Absorption Edge:** Similar to the approach used for Ce, the Pr $L_3$ edge exhibits three distinct peaks labeled A, B, and C, shown in Figure 6a. Peak A corresponds to the $Pr^{3+}$ (2p to $4f^2$) transition, while peaks B and C are associated with $Pr^{4+}$ states, specifically assigned to 2p to $4f^1$ and 2p to $4f^2L$ transitions, where L represents an oxygen ligand hole [45]. To quantify the Pr valence state, the absorption edge was fitted using a series of Lorentzian functions, and the intensity ratio $I_C/I_A$ was used as an indicator of the relative $Pr^{3+}$ fraction, shown in Figure 6b [45]. The estimated valence states for the RE-HEO samples are presented in Table 1. The reference standards $Pr_6O_{11}$ and $Pr_2O_3$, have a known valence of $Pr^{3.667+}$ and $Pr^{3+}$ respectively. The results indicate that the estimated Pr oxidation states for all measured RE-HEO compositions remain consistently between 3.5 and 3.6, with x = 20 and 32.5% closely aligning with the $Pr_6O_{11}$ reference standard. This suggests that Pr in the RE-HEO structure exists as a mixed-valence species, stabilizing in an oxidation state intermediate between $Pr^{3+}$ and $Pr^{4+}$. This implies that Pr maintains a similar local electronic environment and oxidation state regardless of composition variations.

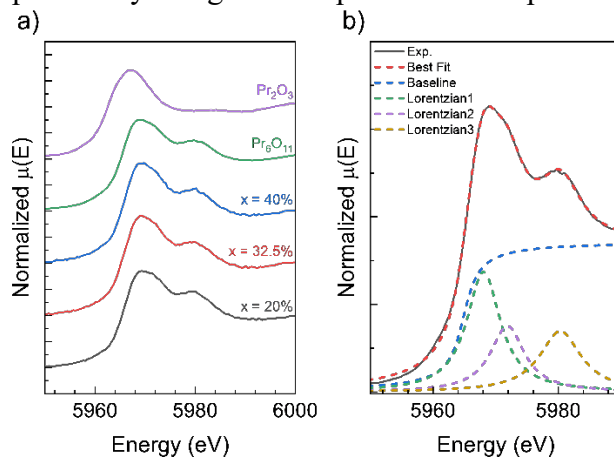

Figure 6: a) Pr $L_3$ absorption of Pr in $Ce_x(SmPrLaY)_{1-x}O_{2-\delta}$ and measured standards and b) an example of fitting procedure performed on $Pr_6O_{11}$.

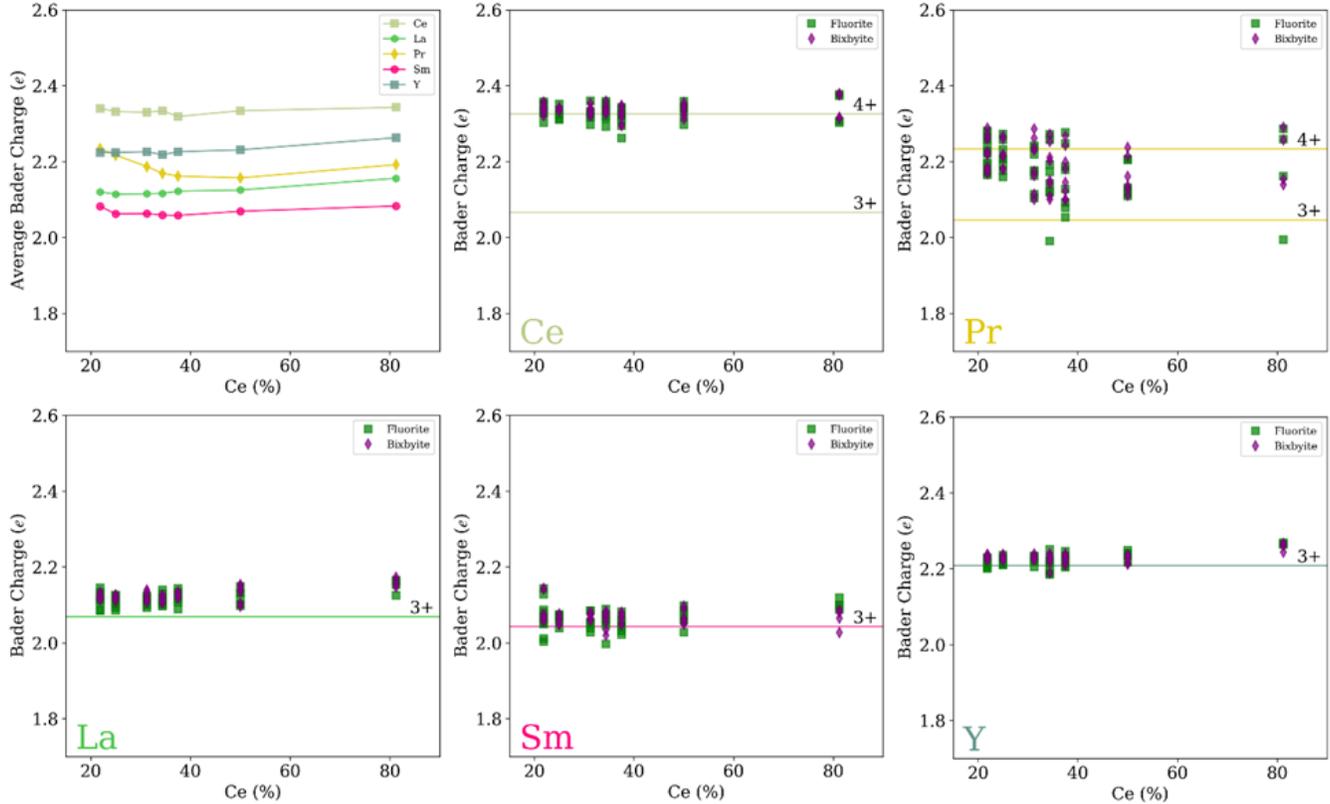

Figure 7: (a) Average Bader charges of Ce, Pr, La, Sm, and Y as a function of Ce concentration in $Ce_x(CeSmPrLaY)_{1-x}O_{2-\delta}$. Lines are included to guide the eye. (b–f) Distributions of individual cation Bader charges for Ce, Pr, La, Sm, and Y, comparing fluorite and bixbyite structures. Horizontal solid lines mark reference Bader charges from the corresponding binary oxides: +3 for La, Sm, Y and both +3/+4 for Ce and Pr.

We also tested the viability of measuring the absorption coefficient of these HEOs in house on an EasyXAFS 300+ (Renton, WA) [46]. The operational details are outlined in supplementary information, along with Supplementary Figure S3 which shows the absorption spectra from in house measurements compared to beamline measurements. XRD and XAFS measurements consistently reveal a Ce-driven structural transition from the bixbyite (Ia-3) to fluorite (Fm-3m) phase across the $Ce_x(SmPrLaY)_{1-x}O_{2-\delta}$ series. At a Ce concentration of 40%, bixbyite-related diffraction peaks disappear, and XANES analysis of the Sm environment indicates a transition to fluorite-like coordination. These results demonstrate that increasing Ce concentration induces both a global symmetry change and a local coordination reorganization. Throughout the series, La and Sm maintain stable trivalent oxidation states. The XANES spectra for La exhibit minimal changes in ligand-field features, suggesting a robust and unperturbed coordination environment, while Sm shows decreasing local disorder at higher Ce concentrations, consistent with the adoption of the more ordered fluorite structure. Ce remains predominantly tetravalent, with a minor, constant $Ce^{3+}$ component, and Pr retains a mixed valence across compositions. XANES analysis indicates that the structural transition is not oxidation state dependent but is instead driven by compositional effects, specifically the increasing fraction of $Ce^{4+}$ and its associated preference for fluorite-type coordination. The relatively consistent $Ce^{3+}$ content suggests that oxygen vacancy concentrations remain largely constant across the series. The persistence of oxygen vacancies may have important implications for tuning ionic conductivity and redox behavior in high-entropy oxide systems.

**DFT Bader Analysis:** Bader charge analysis was performed to assign the total oxidation states associated with each rare-earth cation across a series of $Ce_x(SmPrLaY)_{1-x}O_{2-\delta}$ compositions, spanning both bixbyite and fluorite structures. The atomic charges are subsequently taken as the average of the Bader charges on the rare-earth cations in each oxide compound. Figure 7 shows the average and individual Bader charges for Ce, Pr, La, Sm, and Y as a function of Ce content, with comparisons between fluorite and bixbyite phases. The computed Bader charges are systematically lower than the nominal 3+ or 4+ oxidation states, reflecting the delocalization of electronic charge and the partial covalency of metal–oxygen bonds involving O 2p electrons. This trend is consistent with prior observations [17]. The computed average charges (Figure 4a) reveal that Ce, La, Sm, and Y maintain nearly constant oxidation states across all compositions. With La, Sm, and Y consistent with trivalent character. Among these, Y exhibits the least variation, followed closely by La, while Sm shows slightly broader but still centered distributions near 3+. In contrast, Pr and Ce exhibit more complex behavior. Pr shows a noticeable decrease in average Bader charge between 31% and 38% Ce content before stabilizing, indicating a shift toward a more reduced state in that compositional window (Fig. 8a). This is further supported by the larger distribution of Pr Bader charges in Figure 4c, with many values falling between the 3+ and 4+ reference lines, indicating a mixed valence state. Ce remains largely in the 4+ oxidation state across all compositions, with only a minor population of cations approaching the 3+ reference line (Figure 7b). This result aligns closely with the experimental XANES data, which identified $Ce^{4+}$ as the dominant oxidation state with a small $Ce^{3+}$ component, and supports the mixed-valent nature of Pr.

Comparison between fluorite and bixbyite structures reveals similar charge trends across all cations. Although fluorite supercells show slightly greater variation in Bader charges for some elements, the overall oxidation state distributions remain comparable. This supports the premise that the phase transition observed experimentally—from bixbyite to fluorite with increasing Ce content—does not arise from significant cation redox activity but is instead driven by configurational entropy and the evolution of anion sublattice disorder. Taken together, the DFT-derived Bader analysis reinforces the XANES observation that RE cation valence states are largely invariant with composition. This highlights the important role of oxygen sublattice dynamics, rather than cation redox, in maintaining charge balance across this high-entropy oxide series.

**Conclusions and Further Work:**

This study investigated the impact of Ce concentration on the structural and charge distribution of $(CeSmPrLaY)O_2$ high-entropy oxides. X-ray diffraction (XRD) confirmed a phase transition from bixbyite to fluorite symmetry with increasing Ce content. XAS analysis further revealed the preservation of $La^{3+}$ and $Sm^{3+}$ oxidation states, while Ce maintained a minor $Ce^{3+}$ fraction and Pr exhibited a relatively stable mixed valence. The correlation between cation oxidation states and phase evolution highlights the role of configurational entropy in stabilizing fluorite-like structures. DFT-based Bader charge analysis supports these findings by confirming the stability of most RE cation charge states and revealing only subtle variations in Ce and Pr oxidation behavior across the composition range. Since the cation valences remain consistent across compositions, it is likely that the anion sublattice plays a role in maintaining electroneutrality.

These findings aim to enhance the fundamental understanding of RE-HEOs and their role in phase stability and influence on electronic structure. The ability to decouple oxidation state changes from structural transitions provides an important lens for designing HEOs where valence stability is required. Further studies should explore the role of oxygen vacancies in more detail, particularly how they influence ionic transport and local bonding in the fluorite phase. In addition, temperature-dependent XAS or in situ studies could provide insight into how these materials respond to redox cycling or catalytic environments. Expanding the compositional space to include non-lanthanide or aliovalent dopants may also reveal how charge compensation mechanisms evolve in more chemically diverse high-entropy systems.

**Data availability:**

The data supporting the findings in this study are available from the corresponding author upon reasonable request.

**Author contributions:**

G.R.B. and C.M.R. developed the experimental plan, synthesized all standard compositions, secured beam time, and performed all spectroscopy experiments. S.S.I.A., B.Y., and J-P.M. synthesized high entropy compositions. M.K.C. and S.B.S performed computational analysis. All authors contributed to the writing of results.

**Conflicts of interest:**

The authors declare no competing financial interest.

**Acknowledgements:**

The authors gratefully acknowledge support from the National Science Foundation through the Materials Research Science and Engineering Center DMR 201183. This research used the beamline 12-BM of the Advanced Photon Source, a U.S. Department of Energy (DOE) Office of Science user facility operated for the DOE Office of Science by Argonne National Laboratory under Contract No. DE-AC02-06CH11357.

# Supplementary Information: Lanthanide L-Edge Spectroscopy of High-Entropy Oxides: Insights into Valence and Phase Stability

Gerald R. Bejger[1], Mary Kathleen Caucci[2], Saeed S.I. Almishal[3], Billy Yang[3], Jon-Paul Maria[3], Susan B. Sinnott[3], Christina M. Rost[1]*

1. Department of Materials Science and Engineering, Virginia Polytechnic Institute and State University, Blacksburg, VA 24060, USA.
2. Department of Chemistry, The Pennsylvania State University, University Park, PA 16802, USA.
3. Department of Materials Science and Engineering, The Pennsylvania State University, University Park, PA 16802, USA.
4. Materials Research Institute, The Pennsylvania State University, University Park, PA 16802, USA
5. Institute for Computational and Data Science, The Pennsylvania State University, University Park, PA 16802, USA

*Corresponding author: cmrost@vt.edu

Table S1: Details of the standards used in this work. The standards that required a mixture of powders were milled in a Spex 8000M ball mill and quenched from the tabulated temperature. The space groups were determined from XRD measurements.

| Standard | Space Group | Powder formula | Reaction Temperature | Presumed Absorber Valence | Cation Coordination |
|---|---|---|---|---|---|
| La(OH)$_3$ | P63/m, P-3m | 0.16La(OH)$_3$ + 0.84La$_2$O$_3$ | N/A | La 3+ | 6 |
| LaCrO3 | Pnma | 0.4 La$_2$O$_3$+0.6CeO$_2$ | 1400° C | La 3+ | 9 |
| La$_2$CuO$_4$ | Bmeb | La$_2$O$_3$ + CuO | 1020° C | La 3+ | 8 |
| (La$_{0.4}$Ce$_{0.6}$)O$_2$ | Fm-3m | 0.4La$_2$O$_3$ + 0.6CeO$_2$ | 1400° C | La 3+ | 8 |
| Sm$_2$O$_3$ | C2/m | Sm$_2$O$_3$ | - | Sm 3+ | 7 |
| SmCrO3 | Pnma | Sm$_2$O$_3$ + Cr$_2$O$_3$ | 1300° C | Sm 3+ | 9 |
| Sm$_2$CuO$_4$ | I4/mmm | Sm$_2$O$_3$ + CuO | 950° C | Sm 3+ | 8 |
| (Sm$_{0.2}$Ce$_{0.8}$)O$_2$ | Fm-3m | 0.2Sm$_2$O$_3$ + 0.8CeO$_2$ | 1400° C | Sm 3+ | 8 |
| CeO$_2$ | Fm-3m | CeO$_2$ | N/A | Ce 4+ | 8 |
| CeF$_3$ | P63/mcm | CeF$_3$ | N/A | Ce 3+ | |
| Pr$_6$O$_{11}$ | Fm-3m, R-3 | Pr$_6$O$_{11}$ | N/A | Pr 3.667+ | 6/8 |
| Pr$_2$O$_3$ | I213 | Pr$_2$O$_3$ | N/A | Pr 3+ | 6 |

Table S2: Compositions used in density functional theory simulations.

| x in Ce$_x$(SmPrLaY)$_{1-x}$O$_{2-\delta}$ | δ Values |
|---|---|
| 0.22 | 0.25, 0.28, 0.31, 0.34 |
| 0.25 | 0.28, 0.31, 0.34 |
| 0.31 | 0.25, 0.28, 0.31, 0.34 |
| 0.34 | 0.25, 0.28, 0.31, 0.34 |
| 0.38 | 0.25, 0.28, 0.31, 0.34 |
| 0.50 | 0.22, 0.28, 0.31 |
| 0.81 | 0.22, 0.28, 0.31 |

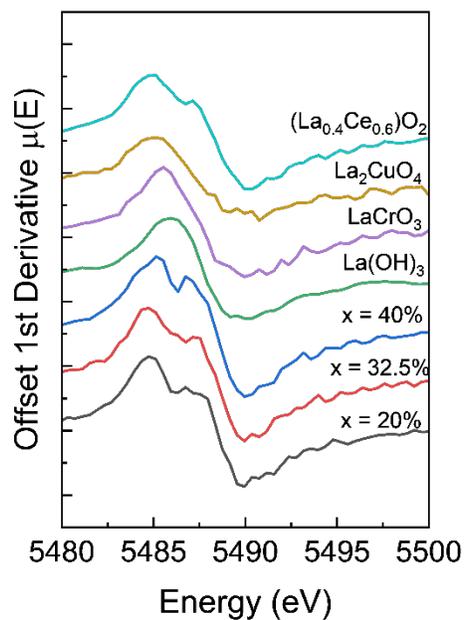

Figure S1: First derivative of La L3 in $Ce_x(SmPrLaY)_{1-x}O_{2-\delta}$ and measured standards.

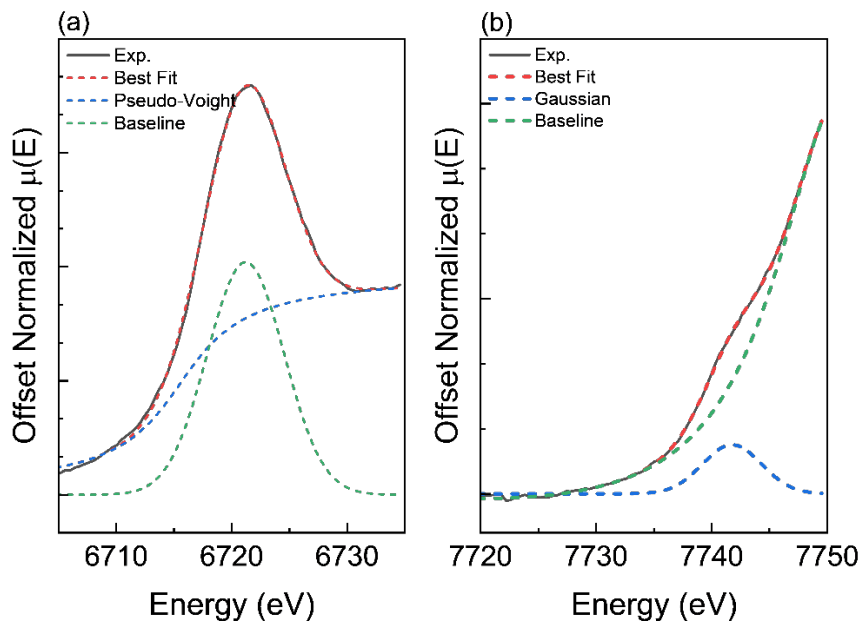

Figure S2: Examples of (a) Sm $L_3$ whiteline fitting and (b) Sm $L_1$ pre-edge peak fitting. The example edges used here are from $Sm_2CuO_4$. Details of fitting procedure is outlined in the text.

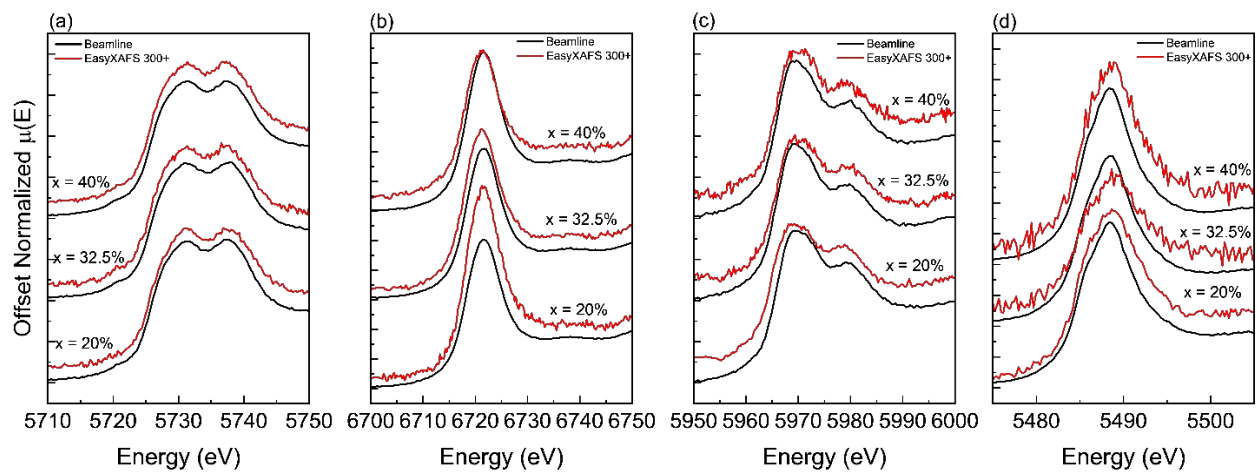

Figure S3: Comparison of L$_3$ absorption edge spectra of (a) La, (b) Sm, (c) Ce, and (d) Pr in $Ce_x(SmPrLaY)_{1-x}O_{2-\delta}$ measured at APS vs on an EasyXAFS300+ operating an Ag X-ray tube with 40 kV and 30 mA using a silicon drift detector.